\documentclass[aps,prb,twocolumn,showpacs,superscriptaddress]{revtex4-1}
\usepackage{graphicx}
\usepackage{multirow}
\usepackage{color}

\begin{document}

\title{Single crystal growth and physical properties of pyroxene CoGeO$_3$}
 
\author{L. Zhao}
\affiliation{Max-Planck Institute for Chemical Physics of Solids, N\"{o}thnitzer str. 40, D-01159 Dresden, Germany}
 
\author{Z. Hu}
\affiliation{Max-Planck Institute for Chemical Physics of Solids, N\"{o}thnitzer str. 40, D-01159 Dresden, Germany}

\author{H. Guo }
\affiliation{Max-Planck Institute for Chemical Physics of Solids, N\"{o}thnitzer str. 40, D-01159 Dresden, Germany}
\affiliation{Neutron Science Platform, Songshan Lake Materials Laboratory, Dongguan, Guangdong 523808, China}

\author{C. Geibel }
\affiliation{Max-Planck Institute for Chemical Physics of Solids, N\"{o}thnitzer str. 40, D-01159 Dresden, Germany}

\author{H. J. Lin}
\affiliation{National Synchrotron Radiation Research Center, 101 Hsin-Ann Road, Hsinchu 30076, Taiwan}

\author{C. T. Chen}
\affiliation{National Synchrotron Radiation Research Center, 101 Hsin-Ann Road, Hsinchu 30076, Taiwan}
 
\author{D. I. Khomskii }
\affiliation{Physics Institute II, University of Cologne, Z\"{u}lpicher Str. 77, 50937 Cologne, Germany}

\author{L. H. Tjeng}\affiliation{Max-Planck Institute for Chemical Physics of Solids, N\"{o}thnitzer str. 40, D-01159 Dresden, Germany}

\author{A. C. Komarek}
\email[]{Alexander.Komarek@cpfs.mpg.de}
\affiliation{Max-Planck Institute for Chemical Physics of Solids, N\"{o}thnitzer str. 40, D-01159 Dresden, Germany}

\date{\today}

\begin{abstract}
We report on the synthesis and physical properties of cm-sized CoGeO$_3$ single crystals
grown in a high pressure mirror furnace at pressures of 80~bar.  
Direction dependent magnetic susceptibility measurements on our single crystals reveal highly anisotropic magnetic properties that we attribute to the impact of strong single ion anisotropy appearing in this system with T$_N$~$\sim$~33.5~K.
Furthermore, we observe effective magnetic moments that are exceeding the spin only values of the Co ions which reveals the presence of sizable orbital moments in CoGeO$_3$.
\end{abstract}
\maketitle

\section{Introduction}

\par
\begin{figure}[!t]
\centering
\includegraphics[width=0.8\columnwidth]{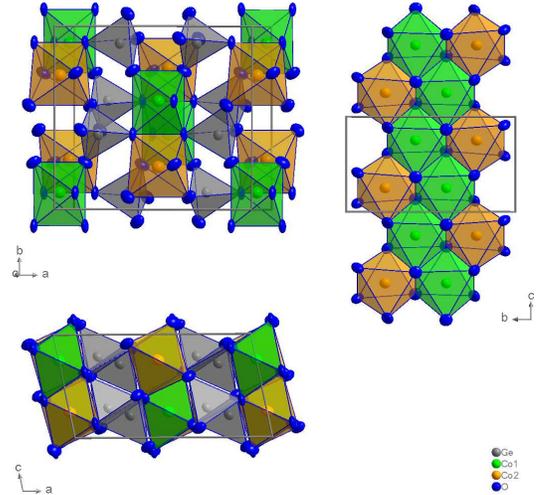}
\caption{ Crystal structure of CoGeO$_3$. The atoms are indicated by 99.9\%\ probability ellipsoids obtained from single crystal X-ray diffraction (see Tabs.~1-2); \textit{grey}: germanium (Ge1), \textit{green}: cobalt (Co1), \textit{dark yellow}: cobalt (Co2) and \textit{blue}: oxygen (O1-O3) atoms. }
\label{crystalstructure}
\end{figure}

 Pyroxenes are one of the main rockforming minerals in the Earth’s crust \cite{geol,geolA,geolB,geolC} and have the general formula $A$$M$$X$$_2$O$_6$ ($A$ = mono- or divalent metal, $M$ = di- or trivalent metal, $X$ = Si$^{4+}$, Ge$^{4+}$ or V$^{5+}$).
This class of materials gained considerable interest due to their large amount of diverse properties \cite{ooA,ooB} including the observation of multiferroicity and magnetoelectric effects  \cite{Jodlauk_2007}.
The quasi 1D system CoGeO$_3$ having two Co sites belongs to the family of pyroxene minerals \cite{Redhammer}
and crystallizes in two polymorphs \cite{Tauber,Redhammer} - the monoclinic polymorph is stable above 1351$^{\circ}$C and the orthorhombic one below this temperature \cite{Tauber}. As reported in literature, single crystals of the monoclinic phase with space group $C2/c$  ($a$ = 9.64 $\AA$, $b$ = 8.99 $\AA$, $c$ = 5.15 $\AA$ and $\beta$ = 101$^{\circ}$10') have been grown from the melt with crystal sizes of the order of 1~$\times$~0.25~$\times$~0.1 mm$^{3}$ \cite{Tauber}. 
The corresponding pyroxene structure of monoclinic CoGeO$_3$ \cite{peacor} consist of Co1 ions that are forming CoO$_6$ octahedral zigzag chains running in $c$-direction, with adjacent Co2 octahedra, compare Fig.~\ref{crystalstructure}. The so formed Co ladders (or double-zig-zag chains) are separated by GeO$_4$ tetrahedra from each other.
Throughout this article we refer to the monoclinic form of CoGeO$_3$ which orders antiferromagnetically below T$_N$~$\sim$~36~K  \cite{Redhammer}.

\section{Results and Discussion}

\begin{figure}[!h]
\centering 
\includegraphics[width=1\columnwidth]{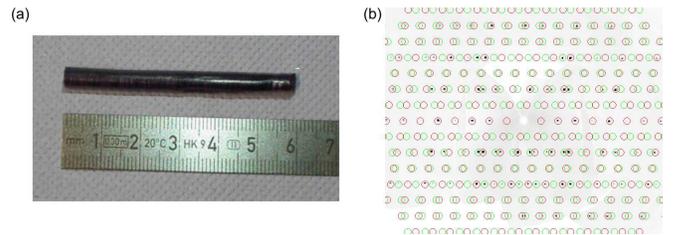}
\caption{(a) Single crystal of CoGeO$_3$ grown in a high pressure mirror furnace. (b) Intensities in the $H$0$L$ plane of reciprocal space measured by means of single crystal X-ray diffraction, compare Tab.~1. These measurements reveal twinning in our single crystals with the twin law matrix (-1 0 -0.732, 0 1 0, 0 0 1). The green and red circles indicate reflections belonging to twin domain A or B.  }
\label{photo}
\end{figure}

\par
\begin{figure}[!b]
\centering 
\includegraphics[width=1\columnwidth]{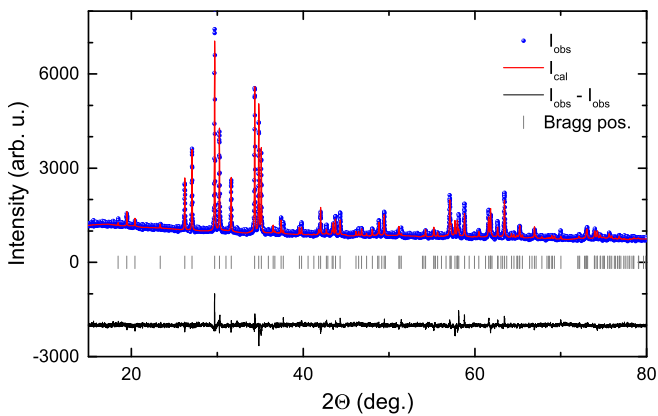}
\caption{(a) Powder X-ray diffraction pattern of a crushed CoGeO$_3$ single crystal.  }
\label{XRD}
\end{figure}

 A photo of our several cm$^{3}$-sized, as-grown single crystal of CoGeO$_3$ is shown in Fig.~\ref{photo}(a). Powder XRD measurements performed on crushed and powderized parts of the single crystal indicate an impurity-free monoclinic phase, compare Fig.~\ref{XRD}.
The lattice parameters obtained from a Rietveld-refinement can be found in the crystal structure table (Tab.~1).
X-ray Laue and single crystal X-ray diffraction measurements indicate the single crystalline nature of our as-grown crystal.
As can be seen in Fig.~\ref{photo}(b) the single crystals are twined with the underlying twin matrix (-1 0 -0.732, 0 1 0, 0 0 1). A precise structural analysis by means of single crystal X-ray diffraction has been performed which confirms that we have synthesized the monoclinic pyroxene CoGeO$_3$ - see Tabs.~1-2.
The resulting positional parameters ($x$, $y$, $z$) for the six different atoms in the asymmetric unit are in agreement with literature data \cite{peacor}, but within the high precision of our measurement we were additionally able to determine the anisotropic displacement parameters - see Tabs.~1-2. From the obtained structural parameters also the accurate bond distances could be determined, see Tab.~3. According to the bond valence sum (BVS) formalism these results indicate Co oxidation states which are very close to 2+, see Tab.~3.

\begin{figure}[!b]
\centering 
\includegraphics[width=0.85\columnwidth]{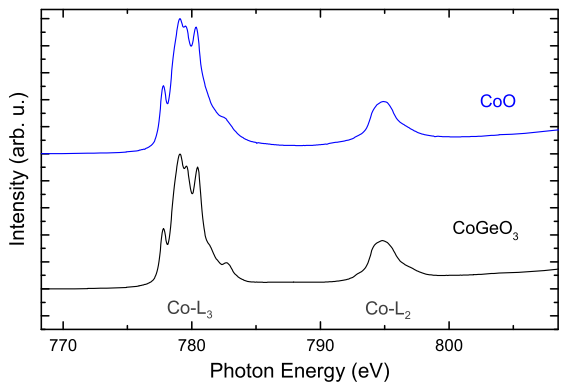} 
\caption{(Isotropic) X-ray absorption spectra  of CoGeO$_3$ and of CoO at the Co L$_{2,3}$ edge.   }
\label{xas}
\end{figure}

The oxidation state of the Co ions in CoGeO$_3$ was further investigated by soft X-ray absorption spectroscopy (XAS) measurements at the Co L$_{2,3}$ edge in the total electron yield mode using a CoO single crystal as a Co$^{2+}$ reference.
The similarity of the Co-L$_{2,3}$ XAS spectra of CoGeO$_3$ and CoO, see Fig.~\ref{xas}, reveals a Co$^{2+}$ high spin state in octahedral coordination \cite{XASa,XASb,XASd} in CoGeO$_3$. These observations corroborate the results of the BVS for the Co ions and further confirm the stoichiometry of our single crystals.

\begin{figure}[!h]
\centering 
\includegraphics[width=0.8\columnwidth]{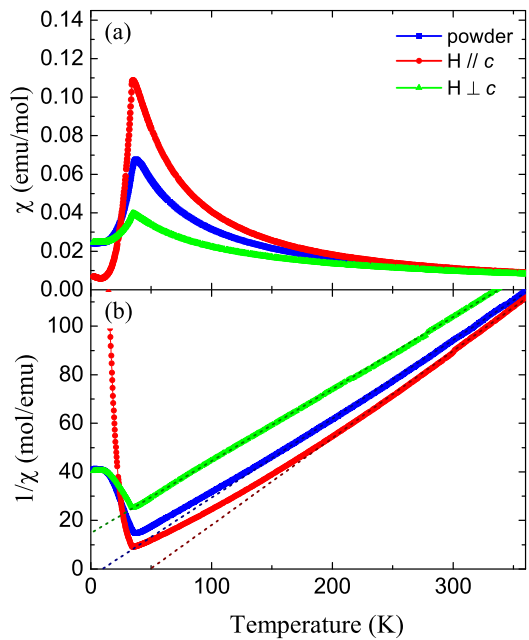}
\caption{(a) Direction dependent magnetic susceptibility ($\chi$) of CoGeO$_3$ single crystals measured in a field of $\mu_0 H$~=~0.1~T. For comparison also the values for a CoGeO$_3$ powder sample are shown. (b) The inverse of the magnetic susceptibility  ($\chi^{-1}$).  }
\label{mag}
\end{figure}

The magnetic susceptibility $\chi$ of CoGeO$_3$ shows a drop at T$_N$~$\sim$~33.5~K, see Fig.~\ref{mag}.
The transition to an antiferromagnetic state is in agreement with literature \cite{Redhammer}.
The availability of sizeable single crystals allowed us to measure also the direction dependence of the magnetic susceptibility.
These direction dependent measurements (with $H || c$ and $H \perp c$) reveal a highly anisotropic behavior
of $\chi$. The Weiss temperatures $\Theta_\mathrm{W}$ obtained from Curie-Weiss fits even have different signs for $H || c$ and $H \perp c$ and amount to  45.08~K and -49.55~K respectively. This strong anisotropy arises from the presence of single ion anisotropy in the system which is typical for Co$^{2+}$-ions \cite{XASa,khomskii}. Moreover, the corresponding effective moments $\mu_\mathrm{eff}$ amount to 4.76~$\mu_{B}$ and 5.18~$\mu_{B}$ respectively. 
The value of the effective moments in CoGeO$_3$ is much larger than the theoretical spin-only value for Co$^{2+}$ ions of 3.87~$\mu_{B}$ and suggests that the Co$^{2+}$ ions are in a high spin state with large orbital moment contributions. 
For powder samples the Weiss temperatures $\Theta_\mathrm{W}$ amount to 6.46~K with effective moments $\mu_\mathrm{eff}$ of 4.98$\mu_{B}$.

\begin{table}
  \begin{tabular}{l|l}
   \hline
Empirical formula   & CoGeO$_3$\\
Formula weight (g/mol) & 179.5 \\
Temperature & room temperature \\
Wavelength & Mo K$_{\alpha}$ \\
Crystal system & monoclinic \\
Space group & \textit{C 2/c} (15) \\
Unit cell dimensions & $a$~=~9.6623(2) \AA \\
                     & $b$~=~8.9928(2) \AA \\ 
										 & $c$~=~5.16980(10) \AA \\
										 & $\beta$~=~101.2785(10)$^{\circ}$ \\
Volume               & 440.535(16) \AA$^3$ \\
Z                    & 8 \\
Density (g/cm$^3$) &  5.4134  \\
Absorption coefficient $\mu$ & 20.861  \\
$F$(000)              & 664 \\
Crystal size  &   $\sim$10-20~$\mu$m  \\
2$\Theta_{max}$ &  106.58$^{\circ}$  \\
Index range   &  $h$: -21~$\rightarrow$~21 \\
              &  $k$: -19~$\rightarrow$~20 \\
							&  $l$: -11~$\rightarrow$~10  \\
Reflections in total / independant & 13219 /  2438  \\
Observed reflections / independant & 10541 / 2078 \\
Internal R-value  &   2.31\%  \\
Completeness up to $2$$\Theta_{max}$ & 91.48\%  \\
Absorption correction & multi-scan \\
Min. / max. transmission &  0.3738 / 0.7505  \\
Refinement method & least squares on $F^2$ \\
Reflections threshold * & $I > 5\sigma(I)$ \\
Goodness of fit &  1.97 \\
R / R$_w$ &  1.63\% / 5.28\%  \\
Largest minima in Fourier difference & -3.10~e$^{-}$~\AA$^{-3}$ \\
Largest maxima in Fourier difference &  2.78~e$^{-}$~\AA$^{-3}$ \\
\hline
 \end{tabular}  
  \caption{Crystallographic \&\ structural refinement data of a single crystal X-ray diffraction measurement.
	The crystallographic software \textit{Jana} was used for the structural refinement \cite{Jana}.
	The lattice parameters were obtained from a powder X-ray diffraction measurement using Cu K$_{\alpha 1}$ radiation (Rietveld refinement with \textit{Fullprof}\cite{fullprof}; $\chi^2$~=~2.24). [*: also unobserved reflections were used.]}\label{xrdZero}
\end{table}

\begin{table}
  \begin{tabular}{l|cccc}
   \hline
	    \textbf{atom}  & \textbf{x} & \textbf{y} & \textbf{z}  \\ 
   \hline
Ge1 &    0.30104(2) &  0.09381(2) &  0.21471(4)  \\ 
Co1 &    0 &  0.09179(4) &  0.75  \\ 
Co2 &    0 &  0.26966(4) &  0.25  \\ 
O1 &    0.11779(15) &  0.09052(14) &  0.1358(3)  \\ 
O2 &    0.38225(14) &  0.24390(16) &  0.3830(3)  \\ 
O3 &    0.36047(15) &  0.06723(16) &  0.9099(3)  \\ 
   \hline   \hline
    \textbf{atom} & \textbf{U$_{11}$ (\AA$^{2}$)} & \textbf{U$_{22}$ (\AA$^{2}$)} & \textbf{U$_{33}$ (\AA$^{2}$)} &       \\
   \hline
Ge1 &  0.00347(12) &  0.00445(13) &  0.00401(12)   \\
Co1 &  0.00526(17) &  0.00499(19) &  0.00458(18)   \\
Co2 &  0.00622(16) &  0.00571(17) &  0.00483(16)   \\
O1  &  0.0020(5)   &  0.0075(6)   &  0.0067(5)   \\
O2  &  0.0074(5)   &  0.0062(5)   &  0.0059(5)   \\
O3  &  0.0077(6)   &  0.0071(5)   &  0.0052(5)   \\
   \hline   \hline
     \textbf{atom} &   \textbf{U$_{12}$ (\AA$^{2}$)} & \textbf{U$_{13}$ (\AA$^{2}$) }& \textbf{U$_{23}$ (\AA$^{2}$)} &    \\
   \hline
Ge1   & -0.00022(5) &  0.00010(8)  & -0.00009(5) \\
Co1   &  0          &  0.00021(13) &  0 \\
Co2   &  0          &  0.00010(12) &  0 \\
O1    &  0.0003(4)  &  0.0003(5)   & -0.0003(4) \\
O2    & -0.0025(4)  & -0.0001(4)   & -0.0012(4) \\
O3    & -0.0019(5)  &  0.0027(5)   & -0.0014(4) \\
   \hline
  \end{tabular}  
  \caption{Refinement results of single crystal X-ray diffraction measurements of CoGeO$_3$.  Our structural results are in fair agreement with older data in literature that contains only isotropic temperature factors $B$ \cite{peacor}. }\label{xrdB}

\end{table}

\begin{table}
  \begin{tabular}{l|lll}
	 \hline
    \textbf{atoms} & \textbf{distance (\AA) / BVS}     \\
   \hline
Ge1-O1 & 1.7399(14)   \\
Ge1-O2 & 1.7142(14)   \\
Ge1-O3 & 1.7965(18)   \\
Ge1-O3 & 1.7963(15)   \\
\textbf{BVS}(Ge1) & \textbf{3.896(8)} \\
\hline
Co1-O1 & 2.0959(15)   \\
Co1-O1 & 2.0959(15)   \\
Co1-O1 & 2.1458(15)   \\
Co1-O1 & 2.1458(15)   \\
Co1-O2 & 2.0649(16)   \\
Co1-O2 & 2.0649(16)   \\
\textbf{BVS}(Co1) & \textbf{1.999(3)} \\
\hline
Co2-O1 & 2.1229(15)   \\
Co2-O1 & 2.1229(15)   \\
Co2-O2 & 2.0157(15)   \\
Co2-O2 & 2.0157(15)   \\
Co2-O3 & 2.2588(17)   \\
Co2-O3 & 2.2588(17)   \\
\textbf{BVS}(Co2) & \textbf{1.894(3)}   \\
   \hline
 \end{tabular}  
  \caption{Bond lengths and bond valence sums (BVS) in CoGeO$_3$. BVS parameters were taken from Ref.~\cite{bvsref}}\label{xrdC}
\end{table}

\section{Materials and Methods}

\par The floating zone growth of monoclinic CoGeO$_3$ was carried out in a high pressure optical mirror furnace (HKZ, \textit{SciDre GmbH}). 
Initially, Co$_3$O$_4$ and GeO$_2$ with an excess of 3\%  GeO$_2$ were mixed together and sintered at 1200$^{\circ}$C for 72 hours with intermediate grindings. From these powders, polycrystalline rods were made using a hydrostatic press and subsequently sintered at 1300$^{\circ}$C for 24~h.

During the floating zone growth pressures of 80~bar of an Argon/O$_2$ mixture (with a ratio of 98 : 2) were used and a growth rate of 3.6~mm per hour was successful for growing large (twined) CoGeO$_3$ single crystals (heating power $\sim$2700~W). 

Powder and single crystal X-ray diffraction (XRD) measurements have been performed on a \emph{Bruker D8 Discover A25} (Cu $K_{\alpha 1}$ radiation) and on a \emph{Bruker D8 VENTURE} diffractometer (Mo $K_{\alpha}$ radiation) respectively.

Soft X-ray absorption spectroscopy measurements have been performed at the BL11A Beamline of National Synchrotron Radiation Research Centre (NSRRC) in Taiwan.
The oxidation state of Co ions in CoGeO$_3$ was determined by soft X-ray absorption spectroscopy (XAS) measurements at the Co L$_{2,3}$ edge in the total electron yield mode using a CoO single crystal as a Co$^{2+}$ reference.

Direction dependent magnetic properties of single crystals of CoGeO$_3$ were initially studied using a SQUID magnetometer (MPMS-5XL, \textit{Quantum Design Inc.}).

\section{Conclusions}

\par We have grown sizable single crystals of CoGeO$_3$ in a high pressure floating zone furnace
that were characterized by XRD and XAS measurements. 
Our direction dependent magnetic susceptibility measurements on our single crystals reveal highly anisotropic magnetic properties with large effective moments of $\sim$5~$\mu_{B}$ per Co ion which are indicative for (i) the significance of single ion anisotropy and (ii) the occurrence of large orbital moments in this system.

\acknowledgments{ 
The research in Dresden was partially supported by the Deutsche Forschungsgemeinschaft through SFB 1143 (Project-Id 247310070). We acknowledge support for the XAS experiments from the Max Planck-POSTECH-Hsinchu Center for Complex Phase Materials. The work of D. Kh. was funded by the Deutsche Forschungsgemeinschaft
(DFG, German Research Foundation) - Project number 277146847 - CRC 1238}

\bibliography{Refs}
 
\end{document}